\documentclass[12pt]{article}
\usepackage{a4wide,amssymb,cite}
\pdfoutput=1

\usepackage{a4wide,amssymb,graphicx}
\usepackage{epsfig}
\usepackage{bm}
\usepackage{here}
\usepackage{cite}
\usepackage[usenames,dvipsnames]{color}
\usepackage{amssymb,cite,graphicx}
\usepackage{slashed}
\usepackage{amsmath,bm,bbm}
\usepackage{amsfonts}
\usepackage[titletoc,title]{appendix}
\usepackage[small]{caption}
\usepackage[margin=1in]{geometry}
\usepackage[multiple]{footmisc}
\usepackage{mathtools}
\usepackage{slashed}
\usepackage[nottoc]{tocbibind}
\usepackage{xcolor}

\newcommand{\be}{\begin{equation}}
\newcommand{\ee}{\end{equation}}
\newcommand{\bea}{\begin{eqnarray}}
\newcommand{\eea}{\end{eqnarray}}

\def\circa#1{\,\raise.3ex\hbox{$#1$\kern-.75em\lower1ex\hbox{$\sim$}}\,}

\begin{document}

\begin{titlepage}
%
%


%

\begin{centering}
\vspace{1cm}
{\Large {\bf A Model of Vector-like Leptons \vspace{.2cm}\\  for the Muon $g-2$  and the $W$ Boson Mass}} \\

\vspace{1.5cm}

{\bf Hyun Min Lee$^\dagger$ and Kimiko Yamashita$^{\star}$  }
\\
\vspace{.5cm}

{\it  Department of Physics, Chung-Ang University, Seoul 06974, Korea.}

\vspace{.5cm}


\end{centering}
\vspace{2cm}

\begin{abstract}
\noindent
We consider a simple extension of the Standard Model (SM) with a vector-like lepton and a local $U(1)'$ symmetry, motivated by the recent experimental anomalies in the muon $g-2$ and the $W$ boson mass.  
The $U(1)'$ symmetry is spontaneously broken  by the VEVs of the dark Higgs scalar and the second Higgs doublet, giving rise to the mixing between the muon and the vector-like lepton.  As a result, we obtain the desirable corrections to the muon $g-2$ and the $W$ boson mass simultaneously, dominantly due to the $Z'$ gauge interactions for the vector-like lepton and the second Higgs doublet, respectively. We also discuss the consistency of the model with the $Z$ boson decay width, the Higgs couplings and the dilepton bounds.

\end{abstract}

\vspace{3cm}

\begin{flushleft} 
$^\dagger$Email: hminlee@cau.ac.kr \\
$^\star$Email: kimikoy@cau.ac.kr
\end{flushleft}

\end{titlepage}

\section{Introduction}

The Standard Model (SM) has been well tested by electroweak precision measurements at collider and intensity experiments and its success has culminated with the discovery of the Higgs boson, being consistent with the predicted couplings in the SM. 
Thus, any new physics scenario beyond the SM must pass the precision tests for electroweak interactions and Higgs couplings, before it becomes meaningful for a further investigation. 

Recently, however, there have been several interesting hints for new physics in some of the precision observables in the SM. 
The long-standing anomaly in the muon $g-2$ observed by Brookhaven E821\cite{Muong-2:2006rrc} has been confirmed by Fermilab E989 \cite{Muong-2:2021ojo}, and the combined significance for the deviation from the SM prediction \cite{Aoyama:2020ynm} is at $4.2\sigma$. Furthermore, another interesting observation for the $W$ boson mass has been very recently made by the CDF II experiment at Fermilab Tevatron \cite{CDF:2022hxs}, with the measured value,
\bea
M^{\rm CDFII}_W= 80.4335\,{\rm GeV}\pm 9.4\,{\rm MeV}. \label{dw}
\eea
This result shows that its deviation from the SM prediction \cite{Haller:2018nnx} is highly significant at the level of $7.0\sigma$, before combination with the previous results from LEP and LHC and the former results in Tevatron experiments \cite{ParticleDataGroup:2020ssz}. 
These two anomalies, the muon $g-2$ and the $W$ boson mass, may call for a simultaneous explanation from a new physics scenario. This is the topic of interest in this article.   
 
We consider an extension of the SM with an $SU(2)_L$ singlet vector-like lepton, which is charged under a local $U(1)'$ symmetry as introduced in Ref.~\cite{Lee:2020wmh,Lee:2021gnw}. We assume that the SM particles are neutral under the $U(1)'$ but the second Higgs doublet carrying the $U(1)'$ charge is introduced for the mixing between the muon and the vector-like lepton. After the electroweak symmetry and the $U(1)'$ symmetry are broken spontaneously, we obtain a small seesaw muon mass and a desirable correction to the muon $g-2$ at the same time. 
Moreover, a nonzero mixing for the left-handed muon modifies the charged current interaction for the muon as well as the $W$ boson self-energy, but it leads to small corrections to the $W$ boson mass, due to the lepton flavor universality for the $Z$ boson decay width. Instead, a nonzero VEV of the second Higgs gives rise to a sizable correction to the $W$ boson mass at the tree level. We show the correlation between the seesaw muon mass, the  muon $g-2$ and the $W$ boson mass in our model.

The paper is organized as follows. 
We first describe the model setup with the vector-like lepton and the $U(1)'$ symmetry. Then, we show how the lepton masses and mixings are generated by a seesaw mechanism in our model and present the modified new interactions for $Z'$ and electroweak gauge bosons. 
Next we discuss the main results of  our work on the corrections to the muon $g-2$ and the $W$ boson.
Finally, conclusions are drawn.

\section{The model}

We introduce an $SU(2)_L$ singlet vector-like lepton, $E$, with charge $-2$ and a dark Higgs field $\phi$ with charge $-2$ under the $U(1)'$  gauge symmetry \cite{Lee:2020wmh,Lee:2021gnw}.
We also have  a Higgs doublet $H'$ with the same quantum numbers as the SM Higgs $H$, except the charge $+2$ under the $U(1)'$. We assume that the SM Higgs doublet $H$ and the SM fermions are neutral under the $U(1)'$. 
The assignments for $U(1)'$ charges are given in Table~1. Here, we chose the $U(1)'$ charges to be $\pm 2$, but if chosen differently, we only have to rescale the $U(1)'$ gauge coupling accordingly, without changing the results.

\begin{table}[hbt!]
  \begin{center}
    \begin{tabular}{c|cccccccccc}
      \hline\hline
      &&&&&&&&&&\\[-2mm]
      & $q_L$ & $u_{R}$  &  $d_{R}$ & $l_{L}$  & $e_{R}$
      & $H$ & $H'$ & $E_L$ & $E_R$ & $\phi$  \\[2mm]
      \hline
      &&&&&&&&&&\\[-2mm]
      $U(1)'$ & $0$ & $0$ & $0$
            & $0$ & $0$ & $0$ & $+2$ & $-2$ & $-2$  & $-2$ \\[2mm]  
      \hline\hline
    \end{tabular}
  \end{center}
    \caption{The $U(1)'$ charges.\label{charges}}
\end{table}

The Lagrangian for the SM Yukawa couplings including $Z'$, dark Higgs $\phi$, and the vector-like lepton is
\bea
{\cal L}=-\frac{1}{4} F'_{\mu\nu} F^{\prime \mu\nu} - \frac{1}{2} \sin\xi F'_{\mu\nu} B^{\mu\nu}+|D_\mu\phi|^2+|D_\mu H'|^2 - V(\phi,H,H') +{\cal L}_{\rm fermions}
\eea
with
\bea
{\cal L}_{\rm fermions}&=& \sum_{i={\rm SM}, E} i {\bar\psi}_i \gamma^\mu D_\mu \psi_i -y_d {\bar q}_L d_R  H- y_u {\bar q}_L u_R {\tilde H} -y_l {\bar l}_Le_R H \nonumber \\
&&  -M_E {\bar E}E-\lambda_E \phi {\bar E}_L  e_R-y_E  {\bar l}_L E_R H'  +{\rm h.c.}. \label{leptonL}
\eea
Here, ${\tilde H}=i\sigma^2 H^*$,  $F'_{\mu\nu}=\partial_\mu Z'_\nu-\partial_\nu Z'_\mu$, $B_{\mu\nu}$ is the field strength tensor for the SM hypercharge, the covariant derivatives are $D_\mu\phi=(\partial_\mu +2i g_{Z'} Z'_\mu)\phi$,  $D_\mu H'=(\partial_\mu -2i g_{Z'} Z'_\mu-\frac{1}{2}i g_Y  B_\mu- \frac{1}{2} i g \tau^i W^i_\mu) H'$,  $D_\mu E=(\partial_\mu+2i g_{Z'} Z'_\mu+i g_Y B_\mu )E$, and $\sin\xi$ is a gauge kinetic mixing.
Moreover,  $V(\phi,H,H')$ is the gauge-invariant scalar potential for the singlet scalar $\phi$, the leptophilic Higgs $H'$ and the SM Higgs $H$, given by
\bea
V(\phi,H, H') &=& \mu^2_1 H^\dagger H + \mu^2_2 H'^\dagger H'  -(\mu_3 \phi H^\dagger H'+{\rm h.c.})  \nonumber \\
&&+ \lambda_1 (H^\dagger H)^2 + \lambda_2 (H'^\dagger H')^2+ \lambda_3 (H^\dagger H)(H'^\dagger H') \nonumber \\
&&+ \mu^2_\phi \phi^*\phi + \lambda_\phi (\phi^*\phi)^2+ \lambda_{H\phi}H^\dagger H\phi^*\phi +  \lambda_{H'\phi}H'^\dagger H'\phi^*\phi \label{scalarpot}
\eea

The $U(1)'$ symmetry and electroweak symmetry are broken by the VEVs of the dark Higgs and the Higgs doublets, $\langle\phi\rangle=v_\phi$, $\langle H\rangle=\frac{1}{\sqrt{2}}v_1$ and $\langle H'\rangle=\frac{1}{\sqrt{2}}v_2$. Then, the squared mass of the $Z'$ gauge boson becomes $m^2_{Z'}=g^2_{Z'}(8 v^2_\phi+ 4 v^2_2)$, before the mass mixing between neutral gauge bosons is taken into account. Moreover, we get the masses for electroweak gauge bosons as $M_Z=\frac{1}{2} \sqrt{g^2+g^2_Y}\,  v$ and $M_W=\frac{1}{2} g v$, with $v=\sqrt{v^2_1+v^2_2}$.

The second Higgs doublet $H'$ couples to the leptons in association with the vector-like lepton, due to the nonzero charge under the $U(1)'$.
Then, the mixing masses between the leptons and the vector-like lepton are generated by the VEVs of the singlet scalar and the leptophilic Higgs doublet, modifying the physical lepton masses and mixings. In order to avoid flavor changing neutral currents at tree level and the collider bounds on light resonances, we assume that the vector-like lepton couples strongly to the muon but very weakly to the other leptons, in particular, to the electron, for instance, to be consistent with the bound from LEP.  For instance, in models with extra dimensions on orbifolds \cite{Asaka:2003iy}, if the SM leptons are localized at different fixed points and the vector-like lepton is localized at the fixed point where the muon is localized, it is possible to explain the muon-specific couplings for the vector-like lepton while the other couplings to electron and tau are suppressed by the volume of the extra dimension.
However, extra vector-like leptons can be introduced to mix with the other leptons in the SM  with no problem for flavor changing neutral currents.

\section{Lepton masses and mixings}

Taking into account the mixing between the vector-like lepton and one lepton, (which is muon for the later discussion on the muon $g-2$), we first enumerate the mass terms for the lepton sector as
\bea
{\cal L}_{L,{\rm mass}}&=& -M_E {\bar E}E-m_0 {\bar e}e-( m_R {\bar E}_L e_R+m_L {\bar e}_L E_R+ {\rm h.c.}) \nonumber \\
&=&- ({\bar e}_L, {\bar E}_L) {\cal M}_L  \left(\begin{array}{c} e_R  \\ E_R \end{array} \right) +{\rm h.c.}
 \label{leptonmass0}
\eea
where 
\bea
 {\cal M}_L= \left(\begin{array}{cc} m_0 & m_L \\ m_R & M_E \end{array} \right), 
\eea
with $m_0$ being the bare lepton mass given by $m_0=\frac{1}{\sqrt{2}} y_l v_1$, and $m_R, m_L$ being the mixing masses, given by $m_R=\lambda_E v_\phi$ and $m_L=\frac{1}{\sqrt{2}} y_E v_2$, respectively.

By making a bi-unitary transformation with the rotation matrices for the right-handed leptons and the left-handed leptons, as follows \cite{Lee:2012wz},
\bea
\left(\begin{array}{c} e_L \\  E_L  \end{array}\right)=U_L\left(\begin{array}{c} l_{1L}\\  l_{2L}  \end{array}\right), \quad
\left(\begin{array}{c} e_R \\  E_R \end{array} \right)=U_R\left(\begin{array}{c} l_{1R}\\  l_{2R} \end{array} \right),
\eea
with
\bea
U_L &=&\left(\begin{array}{cc}  \cos\theta_L & \sin\theta_L\\   -\sin\theta_L & \cos\theta_L \end{array} \right), \\
U_R&=& \left(\begin{array}{cc}  \cos\theta_R & \sin\theta_R\\   -\sin\theta_R & \cos\theta_R \end{array} \right),
\eea
we can diagonalize the mass matrix for the vector-like lepton and the lepton as
\bea
U^\dagger_L {\cal M}_L U_R = \left(\begin{array}{cc}  \lambda_- & 0\\   0 & \lambda_+ \end{array} \right).
\eea
Here, we get the mass eigenvalues for leptons \cite{Lee:2012wz},
\bea
\lambda^2_{-,+}&=& \frac{1}{2} \bigg(m^2_0+ M^2_E +m^2_L +m^{ 2}_R\mp \sqrt{(M^2_E+m^{2}_L-m^2_0-m^2_R )^2+4(m_RM_E+m_L m_0)^2}  \bigg) \nonumber \\
&\equiv& m^2_{l_1,l_2}, \label{masses}
\eea
and the mixing angles as
\bea
\sin(2\theta_R) &=&\frac{2(M_E m_R+m_0 m_L)}{m^2_{l_2}-m^2_{l_1}}, \label{mixR} \\
\sin(2\theta_L) &=&\frac{2(M_E m_L+m_0 m_R)}{m^2_{l_2}-m^2_{l_1}}.   \label{mixL}
\eea

With $\lambda_-=m_{l_1}$ and $\lambda_+=m_{l_2}$ obtained after diagonalization, the lepton masses in eq.~(\ref{leptonmass0}) take  the following form,
\bea
{\cal L}_{L,{\rm mass}}= -m_{l_1} {\bar l}_1 l_1 - m_{l_2} {\bar l}_2 l_2.
\eea
From eq.~(\ref{masses}), we note that the mass eigenvalues satisfy the following simple relation,
\bea
m_{l_1} m_{l_2} = M_E m_0 -m_R m_L.
\eea
Thus, taking $m_{l_2}\simeq M_E$ for $m_0,m_R, m_L\ll M_E$, we obtain the lighter lepton mass as
\bea
\lambda_-=m_{l_1}\approx m_0-\frac{m_R m_L}{M_E}. \label{leptonmass}
\eea
Therefore, the seesaw contribution from the heavy vector-like lepton is naturally small, because it needs
 $m_L\neq 0$ and $m_R\neq 0$, which come from a simultaneous breaking of the electroweak symmetry and the $U(1)'$ symmetry with the second Higgs doublet and the dark Higgs.
As far as $m_0>\frac{m_R m_L}{M_E}$, we find that the lighter lepton mass eigenvalue is positive\footnote{The connection between the seesaw lepton mass and the muon $g-2$ was discussed in Ref.~\cite{Lee:2021gnw}, but the bare lepton mass larger than the seesaw lepton mass is required for a correct sign of the correction to the muon $g-2$.}. Otherwise, in order to get a correct sign for the lepton mass term, we need to make field redefinitions such as $l_{1L}\to l_{1L}$ and $l_{1R}\to -l_{1R}$. For the later discussion, we assume that $m_0>\frac{m_R m_L}{M_E}$ without loss of generality. 

For $M_E\gg m_0 m_L/m_R, m_0 m_R/m_L$, the mixing angles in eqs.~(\ref{mixR}) and (\ref{mixL}) also get simplified to
\bea
\sin(2\theta_R) &\simeq & \frac{2M_E m_R}{m^2_{l_2}-m^2_{l_1}}\simeq \frac{2m_R}{M_E}, \label{mixR2} \\ 
\sin(2\theta_L) &\simeq & \frac{2M_E m_L}{m^2_{l_2}-m^2_{l_1}}\simeq \frac{2m_L}{M_E}. \label{mixL2}
\eea
Then, if $ \frac{m_R m_L}{M_E}\lesssim m_{l_1}$  and $m_R, m_L\ll M_E$, we get $ m_{l_1}\sim \frac{m_R m_L}{M_E}\simeq \theta_R \theta_L M_E$, or $\theta_R\theta_L\sim m_{l_1}/M_E$.

Consequently, the mixing between the lepton and the vector-like lepton leads to the effective interactions of leptons to $Z'$ and weak gauge bosons, as follows,
\bea
{\cal L}_{l,{\rm eff}} &= &-2 g_{Z'} Z'_\mu \Big(c^2_R {\bar E} \gamma^\mu P_R  E+ s^2_R\,  {\bar l} \gamma^\mu P_R l -s_Rc_R ({\bar E}\gamma^\mu P_R l + {\bar l}\gamma^\mu P_R E) \nonumber \\
&&\quad+c^2_L {\bar E} \gamma^\mu  P_L E+s^2_L {\bar l} \gamma^\mu P_L l -s_Lc_L ( {\bar E}\gamma^\mu P_L l + {\bar l}\gamma^\mu P_L E)   
  \Big) \nonumber \\
&&+\frac{g}{2c_W}\, Z_\mu (v_l+a_l) \Big( c^2_L{\bar l} \gamma^\mu P_L l + s_Lc_L( {\bar E}\gamma^\mu P_L l+ {\bar l}\gamma^\mu P_L E)+s^2_L{\bar E} \gamma^\mu P_L E    \Big) \nonumber \\
&&+\frac{g}{2c_W}\, Z_\mu (v_l-a_l) \Big( {\bar l}\gamma^\mu P_R l  + s^2_L {\bar l}\gamma^\mu P_L l - s_Lc_L( {\bar E}\gamma^\mu P_L l+ {\bar l}\gamma^\mu P_L E) \nonumber \\
&&\quad + {\bar E} \gamma^\mu P_R  E+c^2_L {\bar E} \gamma^\mu P_L E \Big) \nonumber \\
&&+\frac{g}{\sqrt{2}}\, W^-_\mu\Big( c_L{\bar l}\gamma^\mu P_L \nu+ s_L {\bar E} \gamma^\mu P_L \nu \Big) +{\rm h.c.},
\eea
with $s_R=\sin\theta_R, c_R=\cos\theta_R$, $s_L=\sin\theta_L, c_L=\cos\theta_L$, and  $v_l=\frac{1}{2}(-1+4s^2_W)$ and $a_l=-\frac{1}{2}$. Here, we used the same notations for mass eigenstates as those for interaction eigenstates, such as $l, E, Z, Z'$.

We remark that the mass mixing between $Z$ and $Z'$ gauge bosons leads to the following extra gauge interactions for the SM fermions and the vector-like lepton, which are separable for small $\theta_{L,R}$, as follows,
\bea
{\cal L}_{Z-{\rm mix}} &=&-2g_{Z'}((c_\zeta-1)Z'_\mu-s_\zeta Z_\mu )({\bar E} \gamma^\mu E) \nonumber \\
&&+ \frac{e}{2c_W s_W}\sum_{f={\rm SM}, E} (s_\zeta Z'_\mu+ (c_\zeta-1) Z_\mu) \Big[{\bar f}\gamma^\mu (\tau^3-2 s^2_W Q_f) f\Big] \label{extraZ}
\eea
where $Q_f$ is the electromagnetic charge operator, $\tau^3$ is the third Pauli matrix, and  $c_\zeta=\cos\zeta, s_\zeta=\sin\zeta$. Here, $\zeta$ is the mixing angle between $Z$ and $Z'$ gauge bosons \cite{Bian:2017xzg,Lee:2021gnw}, given by
\bea
\tan2\zeta= \frac{2m^2_{12} (M^2_{Z_2}-M^2_Z)}{(M^2_{Z_2}-M^2_Z)^2-m^4_{12}} \label{Zmix}
\eea
where $M^2_Z=(g^2+g^2_Y)v^2/4$, 
and the mass eigenvalues for $Z$-like and $Z'$-like neutral gauge bosons are
\bea
M^2_{Z_{1,2}} = \frac{1}{2} \Big(M^2_Z+m^2_{22} \mp \sqrt{(M^2_Z-m^2_{22})^2 +4 m^4_{12}} \Big) \label{Zmasses}
\eea
where
\bea
m^2_{22}&=& m^2_{Z'}/c^2_\xi+M^2_Zs^2_W t_\xi\bigg(t_\xi -\frac{8g_{Z'}}{c_\xi g_Y}\,\sin^2\beta\bigg), \\ 
m^2_{12} &=&M^2_Z s_W t_\xi -\frac{4s_W g_{Z'}}{c_\xi g_Y}\,M^2_Z\sin^2\beta,
\eea
with $\sin\beta=v_2/v$, $t_\xi=\tan\xi$, $c_\xi=\cos\xi$.
Here, we note that the gauge kinetic mixing, $\sin\xi$, also contributes to the mixing between $Z$ and $Z'$ gauge bosons \cite{Bian:2017xzg,Lee:2021gnw}. But, we can ignore the gauge kinetic mixing for $|\tan\xi|\lesssim 4(g_{Z'}/g_Y)\sin^2\beta$. Otherwise, the gauge kinetic mixing should be taken into account for the mass mixing between $Z$ and $Z'$ gauge bosons and the new $Z'$ interactions to the electromagnetic current, being proportional to the gauge kinetic mixing \cite{Bian:2017xzg,Lee:2021gnw}.
As will be discussed in the later section,  the mass mixing between $Z$ and $Z'$ gauge bosons is important for explaining the $W$ boson mass, so it is interesting to consider the effects of the $Z'$ and $Z$ gauge bosons at the LHC, in particular, due to the interactions to the quarks.

We also discuss the SM Higgs Yukawa interactions in the presence of the mixing between the lepton and the vector-like lepton. Considering the mixing angle $\alpha$ between the CP-even neutral Higgs scalars belonging to $H$ and $H'$\cite{Lee:2021gnw} but ignoring the mixing with the dark Higgs coming from $\phi$, we get the effective lepton Yukawa interaction to the SM Higgs $h$, as follows,
\bea
{\cal L}_{h,\mu} &=& -\frac{1}{v} \bigg(\frac{m_0\cos\alpha}{\cos\beta} -\frac{m_L\sin\alpha}{\sin\beta}\,\sin\theta_R \Big) {\bar l}_L l_R h+{\rm h.c.} \label{leptonY}
\eea
where $\cos\beta=v_1/v$. Thus, we take the alignment limit for the extra neutral Higgs with $\alpha=\beta$ and $\sin\theta_R\simeq \frac{m_R}{M_E}$ from eq.~(\ref{mixR2}) for $m_R, m_L\ll M_E $. As a result, from the lighter lepton mass eigenvalue in eq.~(\ref{leptonmass}), the above effective Yukawa interaction for the lepton coincides with the one for the SM. 
Similarly, in the alignment limit, the Higgs Yukawa couplings to quarks are also the same as in the SM \cite{Lee:2021gnw}. 

In passing, we present the symmetry argument for the alignment limit for the Higgs sector with $\alpha=\beta$ in our model. First, we can take vanishing mixing angles between the CP-even scalar coming from the singlet $\phi$ and two CP-even Higgs scalars, for $\lambda_{H\phi}=\mu_3 \tan\beta/(2v_\phi)$ and $\lambda_{H'\phi}=\mu_3 \cot\beta/(2v_\phi)$. Then, we find the squared mass matrix for two CP-even Higgs scalars as
\bea
{\cal M}^2_S= \frac{\mu_3 v_\phi}{s_\beta c_\beta} \left(\begin{array}{cc} s^2_\beta & -s_\beta c_\beta \\ -s_\beta c_\beta & c^2_\beta \end{array} \right)+ v^2 \left(\begin{array}{cc} 2\lambda_1 c^2_\beta & \lambda_3 s_\beta c_\beta \\ \lambda_3 s_\beta c_\beta & 2 \lambda_2 s^2_\beta \end{array} \right).
\eea 
Next, diagonalizing the above mass matrix by the rotation matrix with mixing angle $\alpha=\beta$,  we obtain
\bea
 \left(\begin{array}{cc} c_\beta & s_\beta  \\  -s_\beta  & c_\beta \end{array} \right) {\cal M}^2_S \left(\begin{array}{cc} c_\beta & -s_\beta  \\  s_\beta  & c_\beta \end{array} \right)=  \left(\begin{array}{cc} A & C  \\  C  & B \end{array} \right) 
\eea
with
\bea
A&=&2(\lambda_1 c^4_\beta + \lambda_3 s^2_\beta s^2_\beta + \lambda_2 s^4_\beta) v^2, \\
B&=&\frac{\mu_3 v_\phi}{s_\beta c_\beta}+ 2s^2_\beta c^2_\beta(\lambda_1+\lambda_2-\lambda_3)v^2, \\
C&=& (\lambda_3-2\lambda_1) c^3_\beta s_\beta + (2\lambda_2-\lambda_3) s^3_\beta c_\beta.
\eea
Therefore, we get $C=0$ for arbitrary $\sin\beta$, as far as the Higgs quartic couplings satisfy the following relations,
\bea
\lambda_1= \lambda_2=\frac{\lambda_3}{2}. \label{quarticrel}
\eea 
This case corresponds to $SO(5)\simeq Sp(4)/Z_2$ \cite{Pilaftsis:2016erj} for two Higgs doublets in the conformal sector of the scalar potential in eq.~(\ref{scalarpot}). In other words, if the relations in eq.~(\ref{quarticrel}) or $SO(5)\simeq Sp(4)/Z_2$ symmetry are violated in the Higgs potential, there is no alignment limit, so the effective lepton Yukawa couplings in eq.~(\ref{leptonY}) and the quark Yukawa couplings deviate from those in the Standard Model, being constrained by Higgs data. 

We remark that there are also Yukawa interactions between the lepton and the vector-like lepton via the dark Higgs boson as well as extra Higgs bosons coming from the second Higgs doublet \cite{Lee:2021gnw}. But, we assume that  the extra Higgs scalars coming from the second Higgs doublet are decoupled \cite{Lee:2021gnw}. On the other hand, a nonzero mixing between the dark Higgs and the SM Higgs could modify the Higgs data and introduce a new Higgs decay channel if the dark Higgs is light enough, namely, the SM Higgs decaying into a pair of dark Higgs scalars. 
For a vanishing mixing of the dark and extra Higgs scalars with the SM Higgs boson, we can ignore the effects of the extra scalars in the following discussion on the muon $g-2$ and the $W$ boson mass.

\section{Corrections to muon $g-2$ and $W$-boson mass}

We discuss the results for the corrections to the muon $g-2$ due to the $Z'$ gauge interactions induced by the mixing between the vector-like lepton and the muon. Then, we also show the model prediction for the $W$ boson mass due to the same lepton mixing and consider the consistency of the results with the $Z$ boson decay width.

\subsection{Muon $g-2$}

The combined average for the muon $g-2$ with Fermilab E989 \cite{Muong-2:2021ojo}  and Brookhaven E821 \cite{Muong-2:2006rrc}, the difference from the SM value  becomes
\bea
\Delta a_\mu = a^{\rm exp}_\mu -a^{\rm SM}_\mu =251(59)\times 10^{-11}, \label{amu-recent}
\eea
which shows a $4.2\sigma$  discrepancy from the SM \cite{Aoyama:2020ynm}.

Furthermore, there are measurements for the anomalous magnetic moment of electron \cite{Hanneke:2008tm,Hanneke:2010au} and the experimental measurements in the cases of Cs atoms \cite{Parker:2018vye,Aoyama:2014sxa} and Rb atoms\cite{Morel:2020dww}, which are consistent with the SM prediction, although there is a need of confirmation for the deviation in the fine structure constant measured with the Rb data.

\begin{figure}[!t]
\begin{center}
\includegraphics[width=0.63\textwidth,clip]{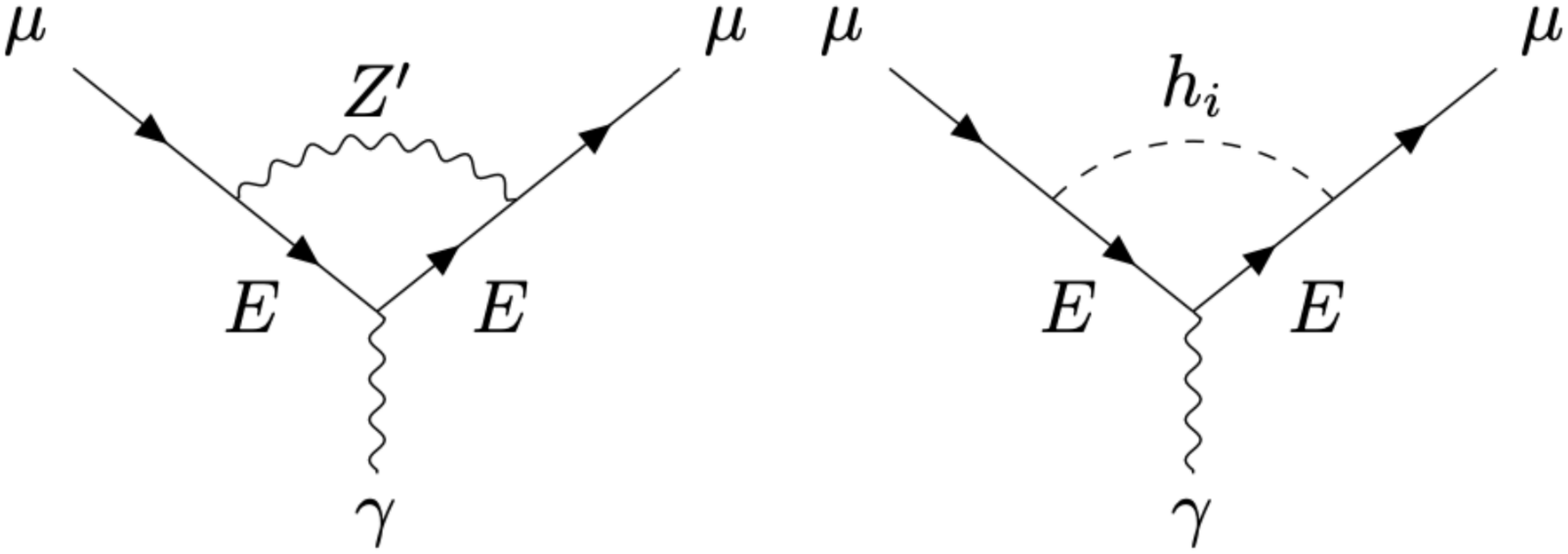} \end{center}
\caption{Feynman diagrams for one-loop corrections to the muon $g-2$, with extra gauge boson $Z'$ on left and extra scalars $h_i$ on right. }
\label{g2Feyn}
\end{figure}

In our model, as shown in the left Feynman diagram in Fig.~\ref{g2Feyn}, the one-loop diagrams with the vector-like lepton and the $Z'$ gauge boson contribute dominantly to the muon $g-2$ \cite{Lee:2021gnw}, as follows,
\bea
\Delta a_\mu&=& \frac{ g^2_{Z'} m^2_\mu}{4\pi^2} \int^1_0 dx \,  \bigg[ c^2_V \Big\{ x(1-x) \Big(x+\frac{2M_E}{m_\mu}-2 \Big)  \nonumber \\
&&-\frac{1}{2m^2_{Z'}} \Big( x^3(M_E-m_\mu)^2 + x^2 (M^2_E-m^2_\mu) \Big(1-\frac{M_E}{m_\mu} \Big) \Big)\bigg\}  +c^2_A \{ M_E\to -M_E\}\bigg] \nonumber \\
&&\times \Big(m^2_\mu x^2 + m^2_{Z'} (1-x) +x(M^2_E-m^2_\mu)\Big)^{-1} \label{amu0}
\eea
where
\bea
c_V&=& \frac{1}{2} (\sin2\theta_R+\sin2\theta_L), \label{cv}  \\
c_A &=& \frac{1}{2} (\sin2\theta_R-\sin2\theta_L). \label{ca}
\eea
On the other hand,  as shown in the right Feynman diagram in Fig.~\ref{g2Feyn}, other one-loop diagrams containing the dark Higgs boson and the extra scalars, $h_i$, with the muon and the vector-like lepton give rise to subdominant contributions to the muon $g-2$  \cite{Lee:2021gnw}.

For a better understanding, we can make an approximation for the corrections to the muon $g-2$ in eq.~(\ref{amu0}) \cite{Lee:2021gnw} as
\bea
\Delta a_\mu \simeq  \left\{\begin{array}{c} \frac{g^2_{Z'} M_E m_\mu}{16 \pi^2 m_{Z'}^2}\, (c_V^2-c_A^2), \qquad M_E\gg m_{Z'}, \vspace{0.3cm} \\ \frac{g^2_{Z'} M_E m_\mu}{4 \pi^2 m_{Z'}^2}\, (c_V^2-c_A^2), \quad m_\mu\ll M_E\ll m_{Z'}. \end{array}  \right. \label{zp1}
\eea
Then, since $c_V>c_A$ from eqs.~(\ref{cv}) and (\ref{ca}), we find  that the correction to the muon $g-2$ is positive, namely, $\Delta a_\mu>0$.
For the small mixing angles for the lepton, we can take $c_V^2-c_A^2\simeq 4\theta_L \theta_R\simeq \frac{4 m_\mu}{M_E}$ from eqs.~(\ref{mixR2}) and (\ref{mixL2}) with seesaw mass $m_\mu\sim m_R m_L/M_E$. We find that there is a non-decoupling effect of the vector-like lepton for the muon $g-2$, because the large chirality-flipping effect from $M_E$ \cite{Crivellin:2021rbq} is cancelled by the small mixing angles such that $\Delta a_\mu\propto g^2_{Z'} m^2_\mu/m^2_{Z'}$, being almost independent of $M_E$ .

\begin{figure}[t]
\centering
\includegraphics[width=0.43\textwidth,clip]{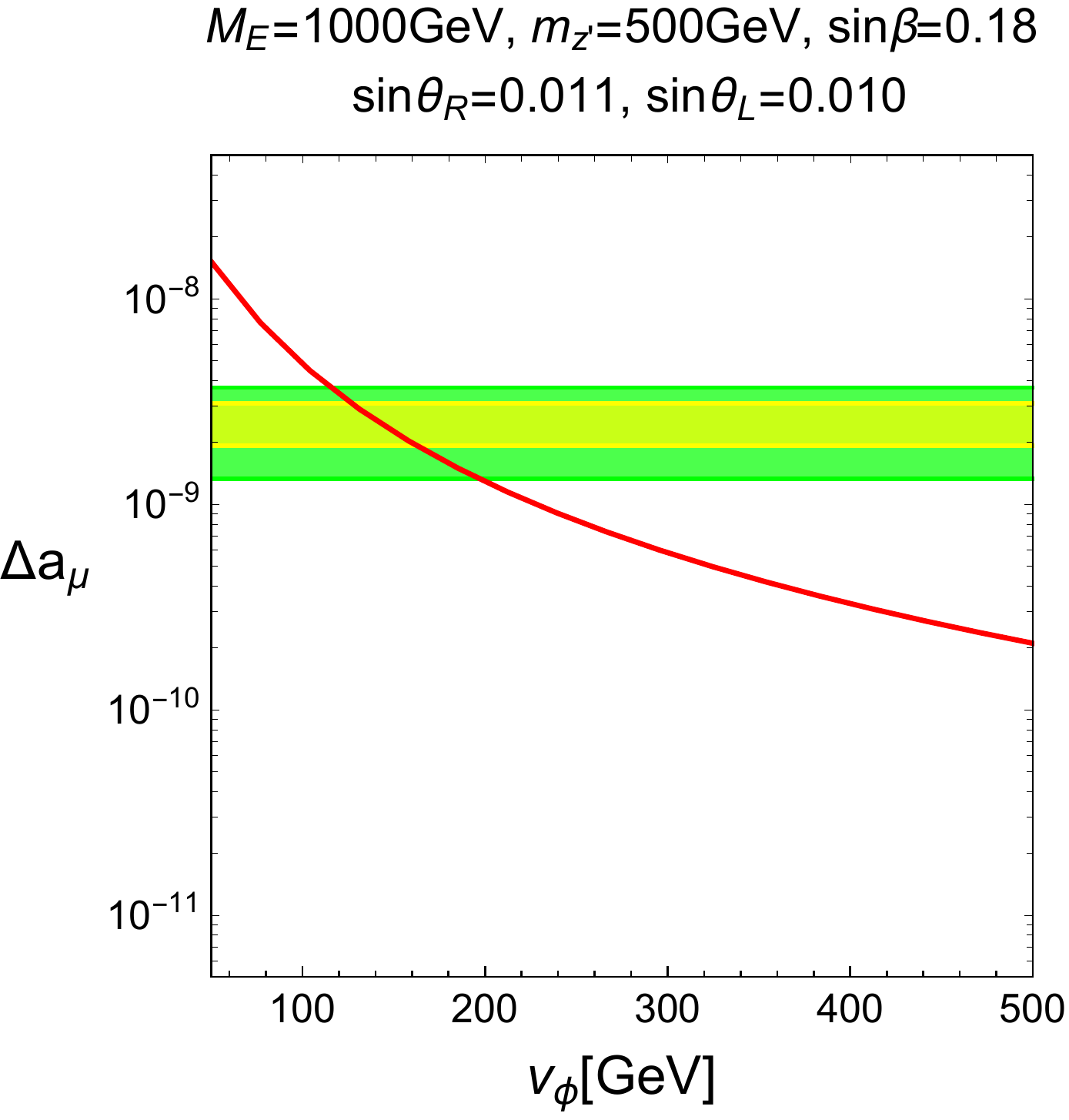}\,\,\,\,\,\,
\includegraphics[width=0.40\textwidth,clip]{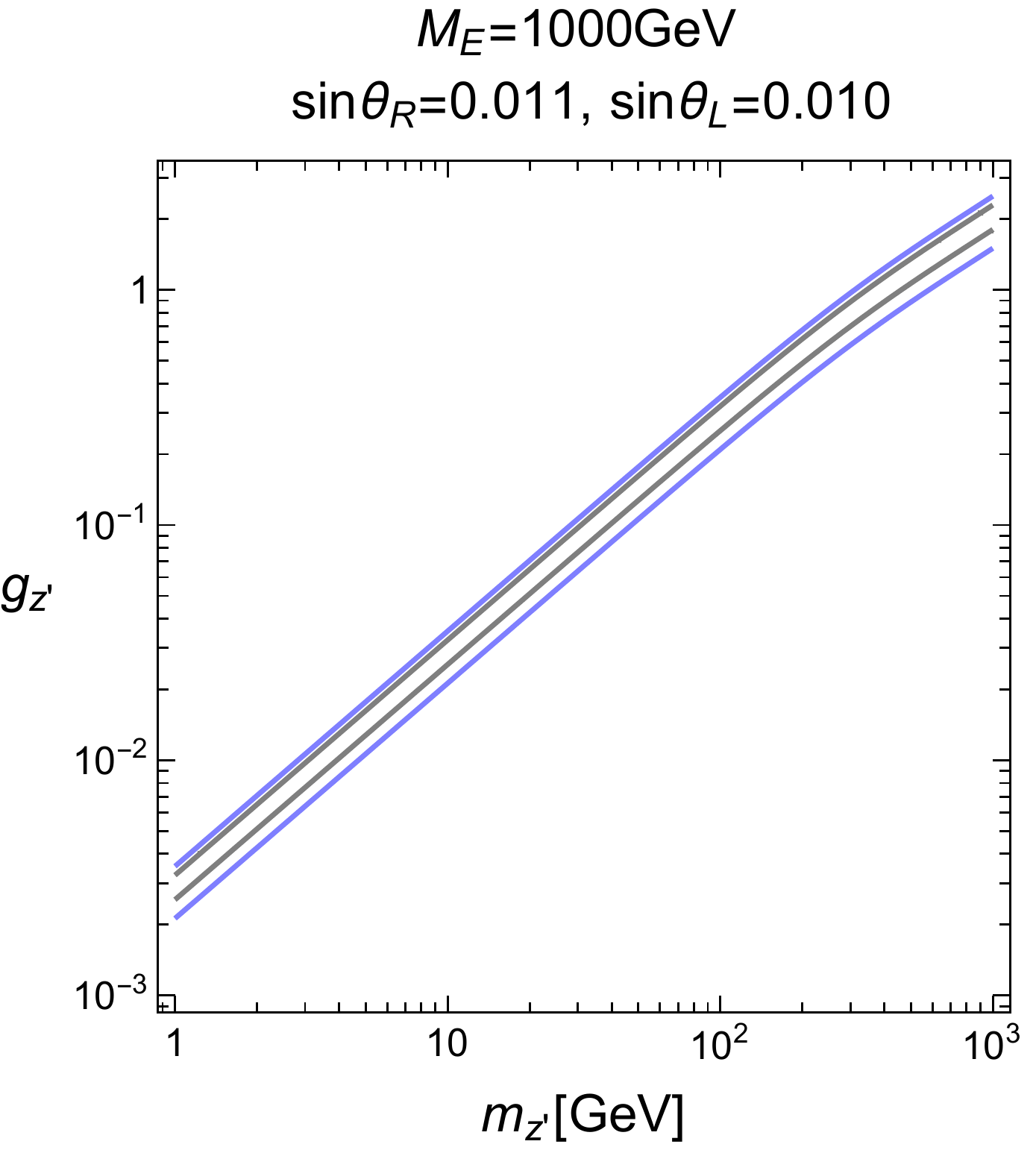}
\caption{(Left) $\Delta a_{\mu}$  as a function of $v_\phi$ in red line, in comparison to $1\sigma$($2\sigma$) bands for the deviation of the muon $g-2$ in yellow(green). We chose $m_{Z'}=500\,{\rm GeV}$ and $\sin\beta=0.18$. (Right) Parameter space in $m_{Z'}$ versus $g_{Z'}$ for explaining  the muon $g-2$ within $1\sigma$($2\sigma$), shown between black(blue) lines. For both plots, we fixed $M_E=1000\,{\rm GeV}$ and  the mixing angles of the vector-like lepton to $\sin\theta_L=0.010$ and $\sin\theta_R=0.011$.  
}
\label{fig:g-2}
\end{figure}

In the left plot of Fig.~\ref{fig:g-2}, we depict the deviation of the muon $g-2$ as a function of the VEV of the dark Higgs $v_\phi$ in our model with $m_{Z'}=500\,{\rm GeV}$. We took  the VEV of the second Higgs to $\sin\beta=0.18$ to get $\Delta M_W=76.5\,{\rm MeV}$ obtained from the Fermilab CDFII experiment \cite{CDF:2022hxs}.
On the other hand, in the right plot of  Fig.~\ref{fig:g-2}, we also showed the parameter space in $m_{Z'}$ versus $g_{Z'}$ for explaining the anomalies of the muon $g-2$ and the $W$ boson mass simultaneously.  The region  between the black(blue) lines in the right plot of  Fig.~\ref{fig:g-2} is consistent with the deviation of the muon $g-2$ at $1\sigma$($2\sigma$) level \cite{Muong-2:2021ojo}.

For both plots in Fig.~\ref{fig:g-2}, we fixed $M_E=1000\,{\rm GeV}$ and took the mixing angles of the vector-like lepton to $\sin\theta_L=0.010$ to be consistent with the lepton flavor universality for the $Z$ boson decay width. Then, we chose $\sin\theta_R=0.011$ for taking the seesaw lepton mass to the muon mass.

\subsection{$W$-boson mass}

The theoretical value of the $W$ boson mass can be derived from the muon decay amplitude, which relates $M_W$ to the Fermi constant $G_\mu$, the fine structure constant $\alpha$, and the $Z$ boson mass $M_Z$\cite{Heinemeyer:2004gx}.
In our model, the charged current for the muon is modified  due to the mixing between the muon and the vector-like lepton, so the $W$ boson mass is determined by the experimental inputs with the following modified formula,
\bea
M^2_W \bigg( 1-\frac{M^2_W}{M^2_Z}\bigg) =\frac{\pi \alpha}{\sqrt{2} G_\mu} \,\cos\theta_L \bigg( 1+ \frac{\Delta r}{\cos\theta_L}\bigg) 
\eea
where $\Delta r$ encodes the loop corrections in the SM and the contributions from new physics.
We note that $\Delta r=0.0381$ in the SM with $\cos\theta_L=1$, which leads to the SM prediction for the $W$ boson mass, as follows \cite{Haller:2018nnx,ParticleDataGroup:2020ssz}, 
\bea
M^{\rm SM}_W=80.357\,{\rm GeV}\pm 6\,{\rm MeV}. 
\eea
On the other hand, the world average for the measured $W$ boson mass in PDG \cite{ParticleDataGroup:2020ssz} is given by 
\bea
M^{\rm PDG}_W= 80.379\,{\rm GeV}\pm 12\,{\rm MeV}.
\eea
Thus, the SM prediction for the $W$ boson mass is consistent with the PDG value within $2\sigma$.
However, according to the Fermilab CDFII experiment  \cite{CDF:2022hxs}, as shown in eq.~(\ref{dw}), the newly measured value for the $W$ boson mass is deviated from the SM prediction at $7.0\sigma$. 

The new physics contribution to the W boson mass can be approximated to
\bea
\Delta M_W \simeq \frac{1}{2} M_W\,\frac{s^2_W}{c^2_W-s^2_W}\, \Big(1-\cos\theta_L-(\Delta r)_{\rm new}\Big), \label{Wmass}
\eea
which is related to the corrections to the $\rho$ parameter by
\bea
(\Delta r)_{\rm new} = -\frac{c^2_W}{s^2_W}\,(\Delta\rho_L+\Delta\rho_H). \label{deltar}
\eea

\begin{figure}[!t]
\begin{center}
\includegraphics[width=0.63\textwidth,clip]{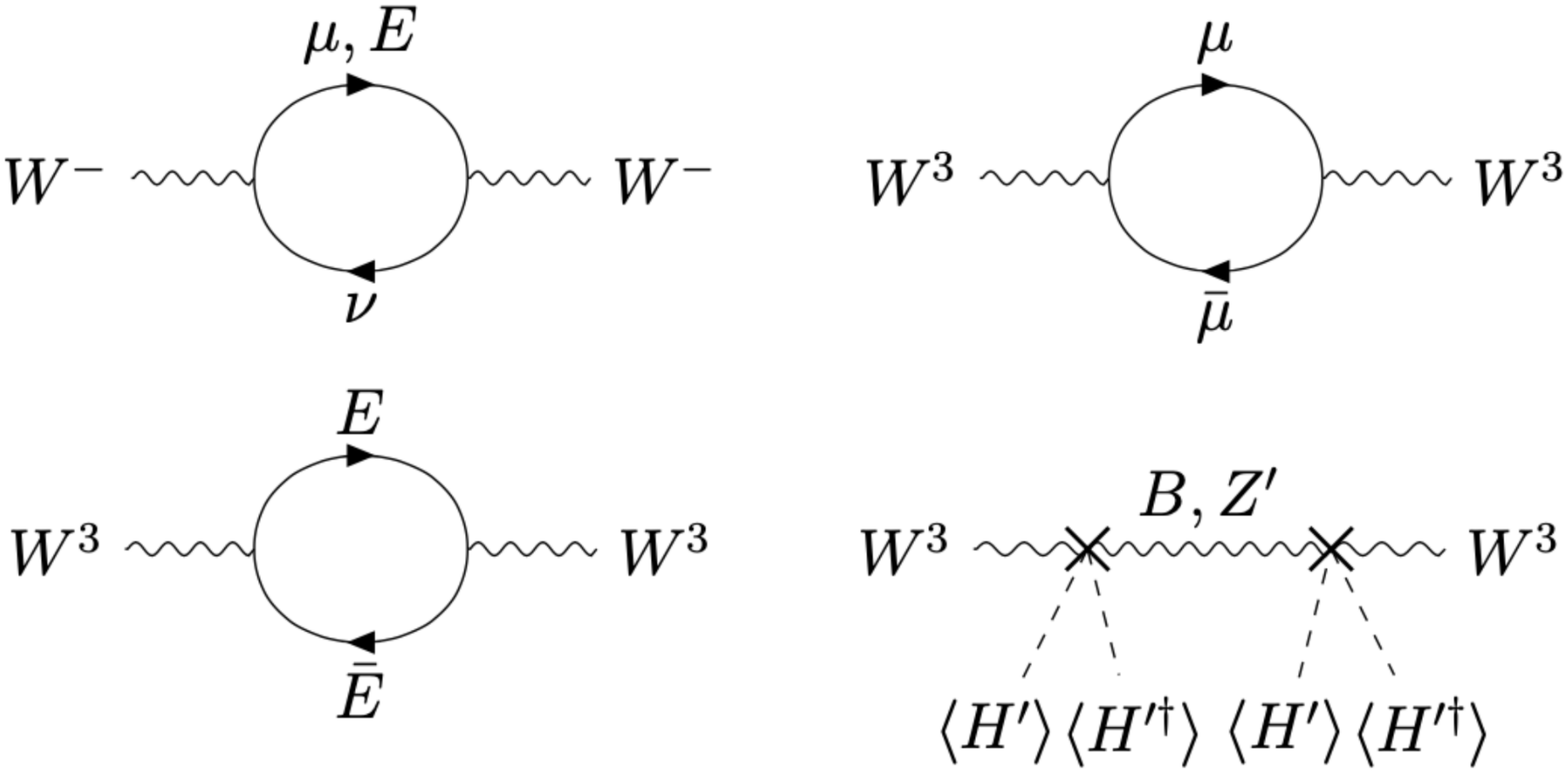} 			
\end{center}
\caption{Feynman diagrams for the self-energy corrections for electroweak gauge bosons.}
\label{WFeyn}
\end{figure}

We now discuss several contributions to the $\rho$ parameter in our model in order.
First, as shown in the first three loop diagrams (two plots in the upper panel and the left plot in the lower panel) in Fig.~\ref{WFeyn}, the $\rho$ parameter is modified by the mixing between the lepton and the vector-like lepton, as follows \cite{Lavoura:1992np,Lee:2021gnw},
\bea
\Delta\rho_L  = \frac{\alpha}{16\pi s^2_W c^2_W}\, \sin^2\theta_L \bigg[\frac{M^2_E}{M^2_Z}-   \frac{m^2_\mu}{M^2_Z}-(\cos^2\theta_L) \theta_+(z_E,z_\mu) \bigg], \label{rhoL}
\eea
with $z_E =M^2_E/M^2_Z$, $z_\mu=m^2_\mu/M^2_Z$, and
\bea
\theta_+(a,b)= a+b -\frac{2ab}{a-b} \ln \frac{a}{b}.
\eea
For $M_E\gtrsim M_Z$, we can make an approximation of the vector-like lepton contribution to the $\rho$ parameter to
\bea
\Delta\rho_L \simeq  \frac{\alpha M^2_E}{16\pi s^2_W c^2_W M^2_Z} \,\sin^4\theta_L,
\eea
which is positive definite, but limited due to the lepton flavor universality from the $Z$ boson width, as will be discussed in the next subsection.

Secondly, as shown in the tree-level diagram (in the right plot in the lower panel) in Fig.~\ref{WFeyn}, the VEV of the second Higgs leads to an additional contribution to the $\rho$ parameter due to the mass mixing between the $Z$ boson and the $Z'$ gauge boson \cite{Bian:2017xzg,Lee:2021gnw}, which is given by
\bea
\Delta \rho_H\simeq  \left\{ \begin{array}{cc} \frac{16s^2_W g^2_{Z'}}{g^2_Y} \frac{M^2_Z}{m^2_{Z'}}\,\sin^4\beta, \quad  m_{Z'}\gg M_{Z}, \vspace{0.3cm} \\  -\frac{16s^2_W g^2_{Z'}}{g^2_Y} \,\sin^4\beta, \quad m_{Z'}\ll M_Z. \end{array} \right.
\eea
Here, we note that a nonzero gauge kinetic mixing, $\sin\xi$, leads to a negative contribution  to the $\rho$ parameter \cite{Bian:2017xzg,Lee:2021gnw}, so we took $\sin\xi=0$ to maximize the new physics contribution to the $\rho$ parameter.
In this case, there is a possibility to increase the $W$ boson mass for $m_{Z'}\gg M_{Z}$, which is crucial to obtain the desirable correction to the $W$ boson mass in our model. 
On the other hand, in the opposite case with $m_{Z'}\ll M_Z$, we can take $\sin\beta\lesssim 0.1\sqrt{g_Y/g_{Z'}}$ to suppress the negative contribution from the second Higgs.    
We note that there is an important difference from the case where the $U(1)'$ charge of the SM Higgs is shifted to a gauge kinetic mixing in Ref.~\cite{Greljo:2022dwn}, because there are two Higgs  doublets with different $U(1)'$ charges in our model.

It is worthwhile to comment on the correlation of the correction to the $\rho$ parameter with the effective electroweak mixing angle for leptons \cite{Wells:2005vk}, as follows,
\bea
\Delta \sin^2\theta^{\rm lept}_{\rm eff} = -\frac{s^2_W c^2_W}{c^2_W- s^2_W}\, \Delta\rho,
\eea
so it is anti-correlated with the correction to the $W$ boson mass.
The world average of the LEP and SLD experiments and the SLD measurement for $\sin^2\theta^{\rm lept}_{\rm eff}$ are given by  $\sin^2\theta^{\rm lept}_{\rm eff}=0.23153 \pm 0.00016$ and $\sin^2\theta^{\rm lept}_{\rm eff}=0.23098\pm 0.00026$, respectively \cite{ALEPH:2005ab}. On the other hand, the SM prediction for $\sin^2\theta^{\rm lept}_{\rm eff}$ is given by $\sin^2\theta^{\rm lept}_{\rm eff}=0.23153\pm 0.00004$ \cite{ParticleDataGroup:2020ssz}. Then, for $\Delta\rho=1.3\times 10^{-3}$ favored by the Fermi CDFII results on the $W$ boson mass, we get $\Delta \sin^2\theta^{\rm lept}_{\rm eff} =-0.00043$, which is in a tension with with the world average at about $2.5\sigma$ but consistent with the SLD measurement. In the following discussion, we don't impose the constraint from $ \sin^2\theta^{\rm lept}_{\rm eff}$ in our model, because the large deviation in the $W$ boson mass with a small uncertainty dominates in determining the parameter space.

\begin{figure}[t]
\centering
\includegraphics[width=0.43\textwidth,clip]{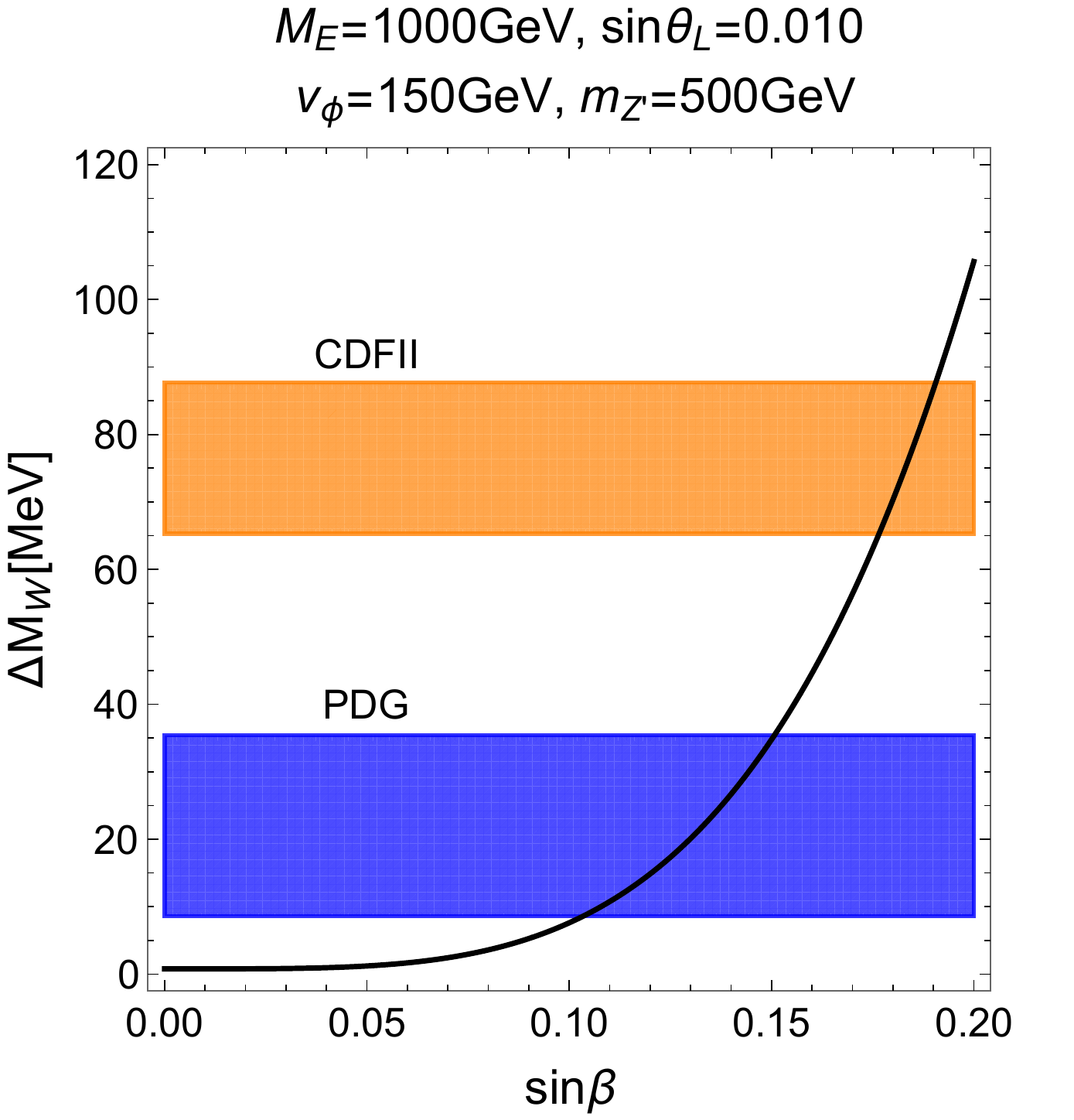} \,\,\,\,\,
\includegraphics[width=0.43\textwidth,clip]{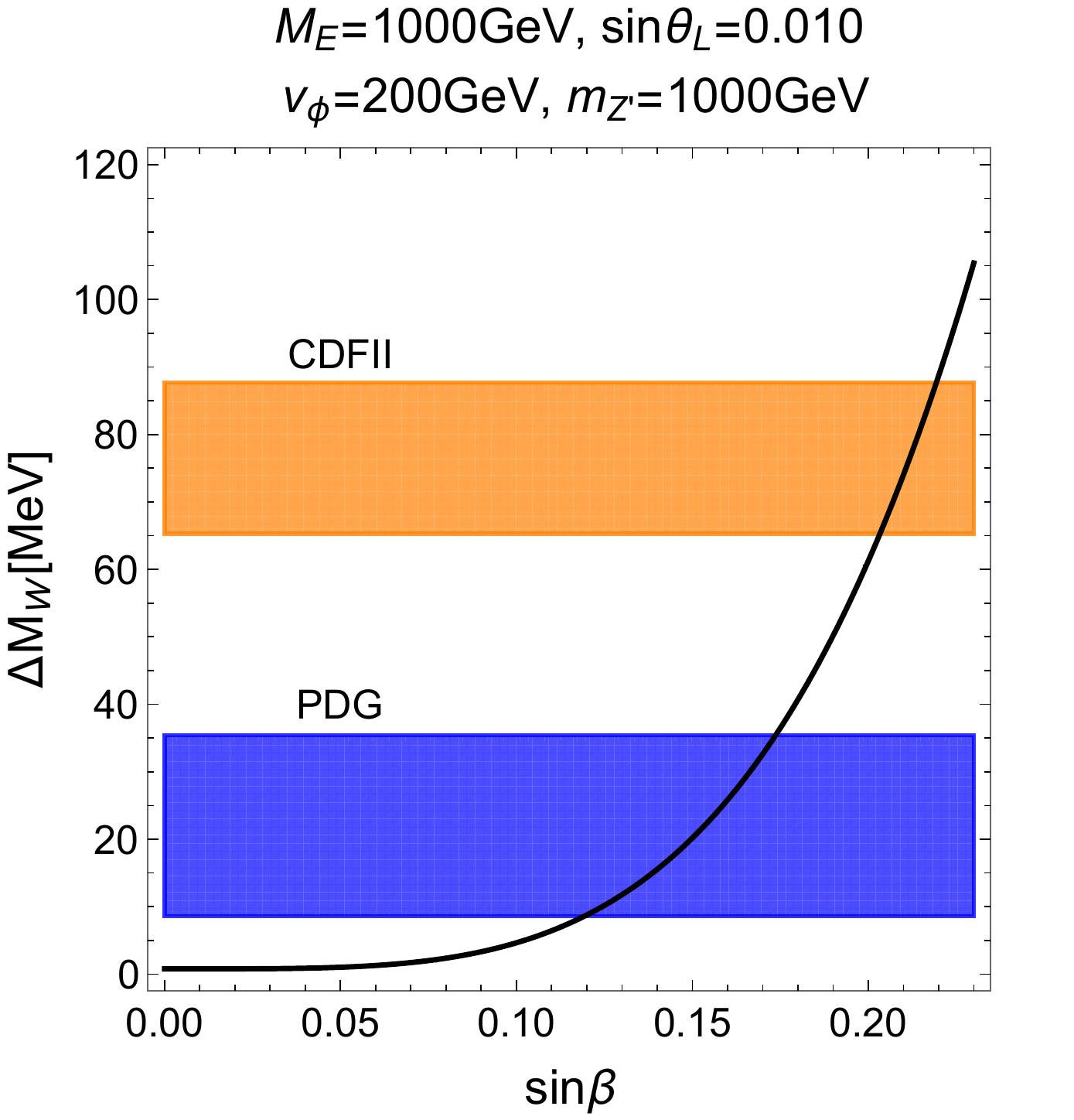}
\caption{$\Delta M_W$  as a function of $\sin\beta$ in black lines.  We took $M_E=1000\,{\rm GeV}$ and $\sin\theta_L=0.010$, in common, and $v_\phi=150(200)\,{\rm GeV}$ $m_{Z'}=500(1000)\,{\rm GeV}$ on left(right). The differences of the $W$ boson mass from the SM prediction in the experimental values obtained from the Fermilab CDFII measurement and the PDG world average are shown within $1\sigma$ errors, in orange and blue regions, respectively.  All the regions are consistent with the $Z$ boson decay width. 
}
\label{fig:mw}
\end{figure}

In Fig.~\ref{fig:mw}, we showed the correction to the $W$ boson mass as a function of $\sin\beta$ in solid line in our model. We compared the deviations from the SM prediction of the $W$ boson mass in the experimental values obtained from the Fermilab CDFII measurement \cite{CDF:2022hxs} and the PDG world average \cite{ParticleDataGroup:2020ssz} within $1\sigma$ errors, as shown in orange and blue regions, respectively.  Together with Fig.~\ref{fig:g-2}, our result shows that  there is a set of the parameters for which the muon $g-2$, the Fermilab CDFII results for the $W$ boson mass and the seesaw lepton mass can be explained simultaneously.  We chose the lepton mixing angle to $\sin\theta_L=0.010$ to be consistent with the bound from the $Z$ boson width.

We comment on the LHC bounds on the mass mixing between the $Z$ boson and the $Z'$ gauge boson, which is important for explaining the $W$ boson mass. In our model, there are also  $Z'$ gauge interactions to the quarks from eq.~(\ref{extraZ}), so the dilepton production from $q{\bar q}\to Z'\to l{\bar l} $  is relevant at LHC. But, since  its cross section scales by $\sin^4\zeta= 5.2\times 10^{-9}$ from eq.~(\ref{Zmix}) for $m_{Z'}=500\, {\rm GeV}$, $v_\phi=150\,{\rm GeV}$ and $\sin\beta=0.2$ for the $W$ boson mass, we can evade the LHC bounds on heavy dilepton resonances \cite{ATLAS:2019erb,CMS:2019tbu}.
Moreover, in our model, up to the contributions from the lepton mixing with $\theta_R\sim \theta_L\sim 0.01$, which are sudominant as compared to those from the $Z$ boson mixing, we obtain the typical effective operator due to the $Z'$ gauge boson, which is relevant for $q{\bar q}\to Z'\to l{\bar l} $, as follows,
\bea
{\cal L}_{qqll} = \frac{C_{u_L}}{v^2} \,({\bar u}_L \gamma^\mu u_L)({\bar e}_L \gamma^\mu e_L)
\eea
with the Wilson coefficient given by
\bea
C_{u_L}= \frac{v^2}{m^2_{Z'}}\Big(\frac{e s_\zeta}{2s_W c_W}\Big)^2(1-2s^2_W)\Big(1-\frac{4}{3} s^2_W \Big). 
\eea
Then, we get  $|C_{u_L}| =3.1\times 10^{-8}$ for  $m_{Z'}=1000 \,{\rm GeV}$, $v_\phi=200\,{\rm GeV}$ and $\sin\beta=0.2$. Therefore, our model is also compatible with the current best bound coming from the high-$p_T$ tail dileptons, $|C_{u_L}|< 10^{-4}$ \cite{Greljo:2017vvb}.

\subsection{Higgs couplings and $Z$ boson width}

First, we remark that the vector-like lepton couples to the SM Higgs boson, modifying the diphoton decay mode of the Higgs boson \cite{Lee:2021gnw}. But, in the alignment limit for two Higgs doublets, it is sufficient to take $\sin\theta_L\lesssim 0.5$ to be compatible with Higgs data \cite{Lee:2021gnw}.

More importantly, the $Z$ boson width can be modified due to the mixing between the lepton and the vector-like lepton.
The partial decay width for $Z\to\mu{\bar\mu}$ in our model is given by
\bea
\Gamma(Z\to \mu{\bar\mu}) = \frac{(v_l+a_l\cos2\theta_L)^2+(v_l-a_l)^2}{(v_l+a_l)^2+(v_l-a_l)^2}\, \Gamma(Z\to \mu{\bar\mu})_{\rm SM} 
\eea
where $\Gamma(Z\to \mu{\bar\mu})_{\rm SM} $ is the partial decay width in the SM. 
The measured total width for the $Z$ boson  \cite{ParticleDataGroup:2020ssz} is given by
$\Gamma_Z=2.4952\pm 0.0023\,{\rm GeV}$, and the ratios of the leptonic branching ratios for the $Z$ boson are measured \cite{ALEPH:2005ab} to be
\bea
\frac{{\rm BR}(Z\to \mu{\bar\mu})}{{\rm BR}(Z\to e{\bar e})}&=&1.0009\pm 0.0028, \\ 
\frac{{\rm BR}(Z\to \tau{\bar\tau})}{{\rm BR}(Z\to e{\bar e})}&=&1.0019\pm 0.0032.
\eea

As a result, we find that the lepton mixing is constrained by the total $Z$ boson width to $\sin\theta_L<0.114 (0.162)$ at $1\sigma(2\sigma)$ level, which is consistent with the desirable correction to the $W$ boson mass. On the other hand, the lepton flavor universality from the $Z$ boson decays constrains the lepton mixing more strongly to $\sin\theta_L<0.030 (0.047)$ at $1\sigma(2\sigma)$ level, for which the contribution of the vector-like lepton  to the $W$ boson mass is limited to $\Delta M_W\lesssim 7.04(17.3)\,{\rm MeV}$ for $M_E\gtrsim 1\,{\rm TeV}$.
We took into account the bounds from the lepton flavor universality for the $W$ boson mass in the previous subsection.

Finally, we comment on a similar test of the lepton flavor universality from the $W$ boson decays. In our model, the leptonic decay, $W\to \mu \nu$ is modified by the lepton mixing angle, $\cos\theta_L$. We first note that ${\rm BR}(W\to\tau\nu)$ measured by LEP was consistently higher than ${\rm BR}(W\to e\nu, \mu\nu)$ by $R_{\tau/\mu}=1.070\pm 0.026$ \cite{ALEPH:2013dgf}, but the LHC data show the consistency with the SM, from $R_{\tau/\mu}=0.992\pm 0.013$ at ATLAS \cite{ATLAS:2020xea} and $R_{\tau/\mu}=0.985\pm 0.020$ at CMS \cite{CMS:2022mhs}. Similarly, the other ratios of the branching ratios for the leptonic decays of the $W$ boson are measured to be $R_{\mu/e}=0.993\pm 0.019$ at LEP, $R_{\mu/e}=1.003\pm 0.010$ and $1.009\pm 0.009$ at ATLAS and CMS, respectively. The experimental uncertainties for the $W$ boson decays are much larger than those for the $Z$ boson decays, so it is sufficient to consider the $Z$ boson decays to constrain the lepton mixing angle as above.

\section{Conclusions}

We presented a simultaneous solution with the vector-like lepton to the anomalies in the muon $g-2$ and the $W$ boson mass, which have recently drawn a lot of attention in particle physics community. The $U(1)'$ symmetry and electroweak symmetry are spontaneously broken by the VEVs of the dark Higgs and the second Higgs doublet, giving rise to nonzero mixing angles between the muon and the vector-like lepton and the tree-level mass mixing between the neutral gauge bosons. Thus, we showed that the small seesaw muon mass and the desirable corrections to the muon $g-2$ and  the $W$ boson mass are generated simultaneously even for a heavy vector-like lepton beyond TeV scale.
A relatively light $Z'$ and a heavy vector-like lepton and the modified neutral and charged currents for the muon can be smoking gun signals in the LHC Run 3 and future experiments.

\section*{Acknowledgments}

We would like to thank Admir Greljo for communication on the $U(1)'$ models and the dilepton bounds.
Moreover, HML appreciates the fruitful discussion with the organizers and the participants in the workshop on the CDFII $W$ boson mass measurement in University of Seoul. 
The work is supported in part by Basic Science Research Program through the National
Research Foundation of Korea (NRF) funded by the Ministry of Education, Science and
Technology (NRF-2022R1A2C2003567 and NRF-2021R1A4A2001897). 
This work of KY is supported by Brain Pool program funded by the Ministry of Science and ICT through the National Research Foundation of Korea(NRF-2021H1D3A2A02038697).

\bibliographystyle{utphys}
\bibliography{ref2.bib}

\end{document}